\documentclass[]{raa}

\usepackage{array}
\usepackage{booktabs}
\usepackage{graphicx, times}
\usepackage{natbib}
\usepackage{amssymb,amsmath}
\usepackage{bm}
\usepackage{xcolor}
\usepackage{siunitx}
\usepackage{subcaption}

\newcommand\rjlr[1]{\textcolor{red} {{\bf #1}}} 
\newcommand\rjlb[1]{\textcolor{blue} {{\bf #1}}}

\newcommand{\atlas}{ATLAS$^{\rm 3D}$}
\begin{document}

\title{Made-to-Measure Outperforms Schwarzschild's Method}


\author{Richard J. Long}


\institute{Department of Astronomy, Westlake University, Hangzhou, Zhejiang 310030, China; {\it rjlastro@yahoo.com; Orcid:0000-0002-8559-0067} \\
           }

\date{Received~~2025 month day; accepted~~2025~~month day}

\renewcommand{\labelitemi}{$\bullet$}
\newcolumntype{M}[1]{>{\centering\arraybackslash}m{#1}}

\abstract{
Syer and Tremaine's made-to-measure method and Schwarzschild's orbit superposition method are well-known within the field of stellar dynamical modeling. 
This research is concerned with assessing and comparing the operational capabilities of the two methods and, in particular, the impact on observable reproduction, orbit classifications and computer elapsed times when using low orbit numbers (8000 orbits) with different observational data sets and initial conditions.
Both methods are able to reproduce observed data with mean $\chi^2 \approx 1$ or less.  However, the made-to-measure process does so three to five times faster than the orbit superposition method, and this starts to make the made-to-measure process attractive for analyzing galaxy surveys.  For a given set of initial conditions, both methods produce similar orbit classifications but the orbits behind the classifications are not the same.  Orbits which are common do not have the same weights.  Different initial conditions result in different classifications. 
}

\authorrunning{Richard J. Long}            
\titlerunning{Made-to-Measure Outperforms Schwarzschild's Method}  

\maketitle


\section{Introduction}
\label{sec:intro}
Syer and Tremaine's made-to-measure method \citep{Syer1996} (M2M) and Schwarzschild's orbit superposition method \citep{Schwarz1979} (SCHW) are well-known within the field of stellar dynamical modeling. Of the two, SCHW has more citations (885) at the time of preparing this article (mid-2025) than M2M (168). M2M and SCHW have been compared previously in \citet{Long2012} where M2M and SCHW were shown to be able to produce similar estimates of galaxy mass-to-light ratios.  It is now appropriate that a further comparison is made.

Three factors are relevant.  
\begin{enumerate}
	\item Data sizes have increased both in terms of the numbers of galaxies being surveyed and the numbers of spaxels being used to capture galaxy data, and will continue to increase. Internally, because of its orbit weighting mechanism, SCHW is not well structured to deal with the increases expected.  M2M is much better positioned.
	\item Galaxy surveys would benefit from being analyzed by methods such as M2M and SCHW with their abilities to handle data from multiple surveys, data of different types (spaxel and discrete), chemical data, population synthesis data and orbit classification constraints.  However both M2M and SCHW appear to be too slow to handle survey scale activities.
	\item An improved understanding of what methods such as M2M and SCHW are capable of doing, or not doing.  For example, M2M and SCHW can only weight the orbits that have been provided to reproduce the observed data.  At best, the models produced may only be illustrative of what the real galaxy might be like.  At worst, the models are non-physical.
\end{enumerate}

This article will not address the third factor: this may be better left to machine learning.  The first two factors will be considered further.  At the time of the first comparison, the accepted assessment was that M2M, with its ability to use significantly more orbits (numbers of $10^5$ to $10^6$ were not uncommon), must of course be producing the better models.  SCHW by comparison was claimed to model real galaxies with only a few hundred orbits, and as a consequence was perceived to be much faster than M2M. M2M, however, with the orbit weights being available from the start of modeling is inherently more flexible than SCHW. The question being posed in this article is how does M2M compare with SCHW when run using fewer orbits, for example $\approx 10^4$.  The objective of this research therefore is to assess and compare the operational capabilities of M2M and SCHW, in particular the impact on observable reproduction, orbit classifications and computer elapsed times, when using low orbit numbers.  This is believed to be the first time that such a comparison between the two methods, under controlled conditions, has been undertaken.

The article's structure is as follows.  Section \ref{sec:approach} summarizes at a top level the approach taken in the investigation.  The observational data are described in Section \ref{sec:galdata}, with methods and techniques in Section \ref{sec:methods}. Results and subsequent discussion are in Sections \ref{sec:results} and \ref{sec:discussion}, with conclusions in Section \ref{sec:conclusions}.  The main abbreviations used are defined in Table \ref{tab:abbrev}.
\begin{table}
	\centering
	\caption{Abbreviations}
	\label{tab:abbrev}
	\begin{tabular}{c|l}
		\hline
		Abbreviation & Meaning \\
		\hline
		3I & three integral initial conditions\\
		\atlas\ & relating to the \atlas\ survey\\
		GPU & graphics processing unit\\
		hmdtu & half mass dynamical time unit (galaxy specific time unit)\\
		IFU & integral field unit\\
		M2M & relating to the made-to-measure method \\
		MDJV & `match density Jeans velocities' initial conditions\\
		MGE & multi-Gaussian expansion\\
		MPI & message passing interface\\
		M/L & mass-to-light\\
		NNLS & non-negative least squares\\
		PC & personal computer\\
		SCHW & relating to Schwarzschild's method\\		
		\hline
	\end{tabular}
\end{table}

\section{Approach} 
\label{sec:approach}
The Introduction, Section \ref{sec:intro}, sets the context and describes \textit{what} is to be investigated. This Approach section describes \textit{how} the investigation is performed.  In this case, the approach is straightforward: it is just a simple combinatorial problem of running a number of models using different methods, initial conditions, and data sets.  The results required for comparison purposes are
\begin{itemize}
	\item the observable mean $\chi^2$ values indicating how well the model has reproduced its observational data,
	\item the computer elapsed time for the modeling runs, and
	\item the orbit classifications giving the weighted hot, warm, cold and counter-rotating orbit fractions supporting reproduction of the observables.
\end{itemize}
The assessment of the performance of M2M relative to SCHW will be based on these comparisons.
Gravitational potentials and initial conditions are specific to the observational data, and are only calculated once since they will be reused whenever the observational data are used.  Data layouts (for example, using a polar grid or Voronoi cells) are not changed except for one set of very specific tests. For other parameters, such as the duration of model runs, their values are kept the same for both the M2M and SCHW runs.

In outline, there are
\begin{itemize}
	\item two methods, M2M and SCHW;
	\item two sets of initial conditions, theoretically based (3I) and observationally based (MDJV); and,
	\item two top level data sets, real galaxy data ( \atlas\ ) and simulated galaxy data (IllustrisTNG).
\end{itemize}
More information is provided in Sections \ref{sec:galdata} and \ref{sec:methods}.  Overall, the comparison has been designed to be between two functionally equivalent methods with implementation and feature usage differences minimized.

It is not intended that the behavior and performance of all the capabilities of M2M will be compared with all the capabilities of SCHW but only the core capabilities concerned with luminosity and kinematical data reproduction.  Features such as discrete data handling, or chemical modeling, or orbit classification constraints are out of scope, as is parameter estimation using the methods.  
Note also that only axisymmetric (oblate) models are utilized: the investigations are not affected by choice of gravitational potential.  The kinematical observables are varied so that both velocity moments and Gauss-Hermite coefficients are modeled - see Section \ref{sec:galdata}.  The earlier comparison in \citet{Long2012} used Gauss-Hermite coefficients only.

SCHW requires the use of NNLS software to create orbit weights,  We choose NNLS software that implements the Lawson and Hanson active set method \citep{LH1974} since it appears from the literature to be the most used NNLS implementation for SCHW. The active set algorithm deliberately zeroises the weights of any orbits it chooses not to use in reproducing the observed data, and this is taken into account in the analysis software used.

Model selection frameworks such as \cite{Lipka2021} and \cite{Thomas2022} are not used.  Some use is made of an experimental Gaussian processes approach to M2M hyper-parameter determination.

Other than to compare M2M and SCHW orbit classifications for IllustrisTNG data with \cite{Xu2019}, no external comparisons of models are made.

\section{Observational Data}
\label{sec:galdata}
The observational data sets used in this article are the same as those used in \citet{Long2025}.  The data tables here are modified versions of the tables from that paper.

For the \atlas\ data set \citep{AtlasI}, data from four galaxies are utilized - NGC 1248, NGC 3838, NGC 4452, and NGC 4551.  These galaxies have various inclinations, different mass-to-light ratios, and sense of bulk rotation (see Table \ref{tab:atlasdata}).  Surface brightness values are calculated using multi-Gaussian expansions (Sect. \ref{sec:gravpot}).  The IFU cells for kinematical data (first and second velocity moments) are the Voronoi cells resulting from signal-to-noise processing of the raw data.
 
\begin{table}
	\centering
	\caption{Top level data for the \atlas\ galaxies}
	\label{tab:atlasdata}
	\begin{tabular}{ccccccc}
		\hline
		Galaxy & Inclination & M/L ratio & Type & Bulk rotation & SB max radius & IFU cells\\
		\hline
		NGC 1248 & $42 ^{\circ}$ & $2.50$ & S0 &  Counter clockwise & 3.00 & 297\\
		NGC 3838 & $79 ^{\circ}$ & $4.00$ & S0 &  Clockwise & 5.86 & 383\\
		NGC 4452 & $88 ^{\circ}$ & $5.20$ & S0 &  Clockwise & 5.17 & 489\\
		NGC 4551 & $63 ^{\circ}$ & $4.89$ & E  &  Counter clockwise & 3.41 & 596\\
		\hline
	\end{tabular}
	
\medskip
The inclinations are taken from \citet{AtlasXV}, while the M/L ratios were determined by M2M modeling in \cite{Long2016}.  The SB max radius column gives the overall size of the polar grid used for surface brightness data values.  The IFU column gives the number of Voronoi cells for the kinematical data. 
\end{table}

For the IllustrisTNG galaxies \citep{Nelson2019}, five axisymmetric galaxies (A0490, A1090, A1190, A1290, and A1390) are utilized, and are viewed edge on with the mass-to-light ratio taken as $1.0$ for all galaxies.  Surface brightness values are again calculated using multi-Gaussian expansions. The IFU cells for kinematical data (the Gauss-Hermite coefficients $h_1$ to $h_4$) are constructed to represent a real galaxy's data (see Table \ref{tab:simdata}).
 
\begin{table}
	\centering
	\caption{Top level data for the simulated galaxies}
	\label{tab:simdata}
	\begin{tabular}{ccccccc}
		\hline
		Galaxy & Bulk rotation & SB max radius & IFU cells\\
		\hline
		A0490 & Counter clockwise & 7.00 & 220\\
		A1090 & Counter clockwise & 6.00 & 197\\
		A1190 & Clockwise         & 6.00 & 110\\
		A1290 & Clockwise         & 6.00 & 173\\
		A1390 & Clockwise         & 7.00 & 119\\
		\hline
	\end{tabular}
	
\medskip
As in Table \ref{tab:atlasdata}, the SB max radius column gives the overall size of the surface brightness polar grid, and the IFU column gives the number of Voronoi cells.  Inclinations for all the simulated galaxies are $90 ^{\circ}$, and mass-to-light ratios are $1.0$.
\end{table}

For both data sets, data values are symmetrized appropriately for an axisymmetric model/potential.  The units used are effective radii for length, $10^7$ years for time, and mass in units of the solar mass with luminosity similarly so.

\section{Methods} 
\label{sec:methods}
With the exception of M2M, all the methods and techniques in this section are very similar to those in \citet{Long2025} and are not described in detail here. The M2M implementation was used recently in \cite{Long2021}, and is described more fully in \citet{Long2016} and other papers by the lead author prior to that.

\subsection{Gravitational Potentials}
\label{sec:gravpot}

The gravitational potentials and their derivatives used for modeling are derived from MGEs of the surface brightness data \citep{Emsellem1994, Cappellari2002}.  For the IllustrisTNG simulated data, dark matter potentials are included as spherical Gaussians, and no black hole modeling takes place.  For the \atlas\ data, no dark matter is included for consistency with \cite{AtlasXV}, and central black holes are modeled as point sources.

\subsection{Orbit Initial Conditions}
\label{sec:ics}

Initial phase space coordinates for stellar orbits are created using the 3I scheme (theoretically based) and the MDJV scheme (observationally based).  These initial coordinates are given an overall sense of rotation matching that shown by the observed mean line-of-sight velocity data, with 75\% of the orbits having the same sense of rotation and 25\% counter-rotating. In general, $8000$ orbits are used for most of the modeling of the observed data sets.  For \atlas\ galaxies NGC 3838 and NGC 4452, $16000$ orbits are used instead so that the elapsed time impact on the SCHW NNLS process, in particular, may be assessed.

\subsection{Orbit Classifications}
\label{sec:orbittypes}

The circularity measure $\lambda _z$ \citep{Zhu2018} is used to classify orbits into one of hot ($\left|\lambda _z\right| <= 0.25$), cold ($\lambda _ z >= 0.80$), warm ($0.25 < \lambda _z < 0.8$) or counter-rotating. The total orbit weights in each of the classes are taken as the overall orbit classification for a galaxy.  Comparing orbit classifications produced by M2M and SCHW is a key part of our modeling analysis.  Note that orbit classifications are not used as constraints themselves, as in \citet{Long2025}.  However, the sum of the weights is constrained to be equal to one, and so the sum of the orbit classes must also equal one.

\subsection{SCHW Models}
\label{sec:schw}
SCHW has three main phases in its execution.
\begin{description}
	\item[\textit{Phase 1}] Using the initial conditions as orbit start coordinates, orbit trajectories are created in the gravitational potential that has been associated with the observed data.  The phase space coordinates representing the trajectories are used to construct the individual orbits' contributions to the model equivalents of the observed data.  \textit{Phase 1} completes once a predetermined number of orbit integration time steps or orbits have been completed.
	\item[\textit{Phase 2}] The individual orbit contributions are weighted in a non-negative least squares sense (NNLS) such that the model observables reproduce as closely as possible the observed data.  Any regularization constraints are included in this phase.
	\item[\textit{Phase 3}] The weighted orbits are analyzed as necessary to learn more about the observed data. 
\end{description}

For use with SCHW, the observed data must be modified so that it is surface luminosity (or mass) weighted.  In addition, for this investigation, regularization is used (heavy weights penalization) in order to make realistic comparisons with M2M possible (see Sections \ref{sec:m2m} and \ref{sec:regSCHW}).

No use of SCHW dithering (orbit bundling) is made as it causes a significant performance overhead for SCHW.

\subsection{M2M Models}
\label{sec:m2m}
M2M differs from SCHW in that \textit{phase 1} and \textit{phase 2} are combined into a single phase in which the weights exist from the start of modeling and are iteratively adapted, as the orbit trajectories are traversed, such that the model observables (calculated using the weights) eventually reproduce the observed data.  This iterative adaption of the weights is akin to the back-propagation mechanism in machine learning neural networks.  M2M terminates in a similar fashion to SCHW phase 1.  On termination, an extra condition must be met which is that the weights should have converged to constant values (to within some numerical tolerance).

Exponential smoothing is used by default in the construction of M2M model observables.  This smoothing mechanism could be turned off but, for the low number of orbits used here, results in much noise within the model, for example in the observable $\chi^2$ series.  The preferred approach here is to regularize SCHW instead (see Sections \ref{sec:m2m} and \ref{sec:regSCHW}) to achieve an approximate balance in behavior of the two methods.

\subsection{Hyper-parameters using Gaussian Processes}
\label{sec:gaussproc}
Both modeling methods require a number of hyper-parameters to be set, for example, the $\epsilon$ learning rate parameter in M2M, or the regularization parameter in SCHW.  In general, these parameters have been set by a combination of experience and experimentation.  During this investigation, an experimental routine using Gaussian processes was developed using the \textit{scikit-optimize} package's \textit{gp\_minimize} interface.  This routine was used to good effect to confirm some of the M2M $\epsilon$ settings, and also to demonstrate that M2M can be used successfully inside a more general framework application.

\subsection{Software Utilization}
\label{sec:computil}

As in \cite{Long2025}, the software base for constructing initial conditions and for M2M and SCHW modeling is the lead author's implementation of the \cite{Syer1996} M2M stellar dynamical modeling method together with the \cite{Schwarz1979} orbit superposition method (a single code set supports both methods).  This software was first used in \cite{Long2010}, and, as indicated, more recently in \cite{Long2025}. The software is Python 3 based with some use of Cython \citep{behnel2010cython} and C for performance critical code. 
The MPI message passing interface is used to achieve parallelization of orbit integration, with the same number of cores being used for both M2M and SCHW.  The optimization software used for NNLS is taken from the SCIPY package version 1.16  \citep{2020SciPy-NMeth} (earlier versions may have NNLS performance issues). The Gaussian processes application is based on routines within the publicly available \textit{scikit-optimize} package.  Using a $20$ core desktop PC, multi-processor working is employed with up to 10 cores active in a modeling run.

\section{Results}
\label{sec:results}
Most of the column headings in the tables within this section should be self-explanatory. However, for clarity,
\begin{itemize}
	\item `Active orbits' refers to the number of orbits that have a non-zero weight while `Orbits' indicates the number of orbits in the initial conditions.
	\item Under `Mean $\chi^2$ values', `sb  los v1 los v2' indicates surface brightness and the line-of-sight first and second velocity moments.  `sb $h_1$ $h_2$ $h_3$ $h_4$' indicates surface brightness and the Gauss-Hermite coefficients $h_1$ to $h_4$.
	\item Under `Orbit classification', `Hot Warm Cold C-R' refers to the hot, warm, cold and counter-rotating class values described in Section \ref{sec:orbittypes}.
\end{itemize}

As an aid to the reader, tables used in comparing orbit classification results employ colored text to indicate differences of $\geq 3\%$ (as in \citealt{Long2025}).  What is important in these tables is to be clear on the direction the comparison is being made (horizontally or vertically between sections in the tables), and to use the colors to gain a visual impression of the extent of any disagreement.  For example, a table section with few colored entries means few differences implying results are similar; conversely, many colored entries imply that results are dissimilar.  Note that the sum of the orbit classes in a classification for a galaxy, as stated in Section \ref{sec:orbittypes}, must be one.  This means that an increase or decrease in one class value must result in decreases or increases in one or more other class values.

\subsection{Regularizing SCHW}
\label{sec:regSCHW}

\begin{figure}[h]
	\centering
	\caption{Heavy Weights Penalization illustrated with NGC 4452}
	\label{fig:heavypen4452}
	\begin{tabular}{ccc}
		SCHW with no penalization & SCHW penalized & M2M \\
		\includegraphics[width=50mm]{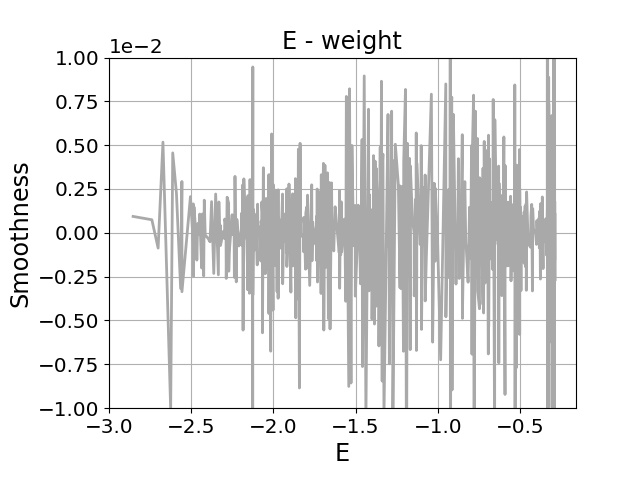}  & \includegraphics[width=50mm]{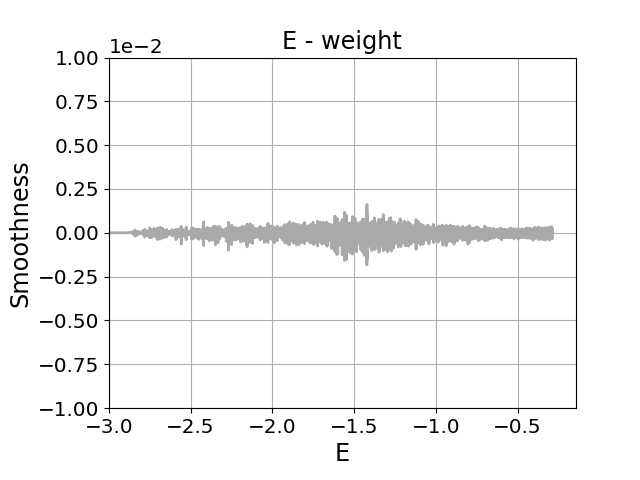}   & \includegraphics[width=50mm]{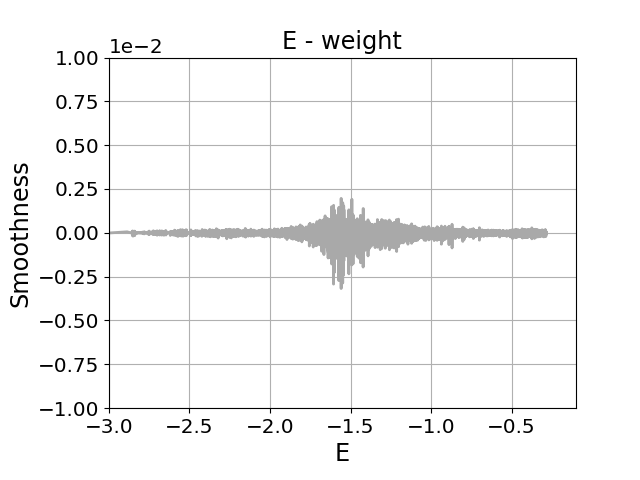}\\
		\includegraphics[width=50mm]{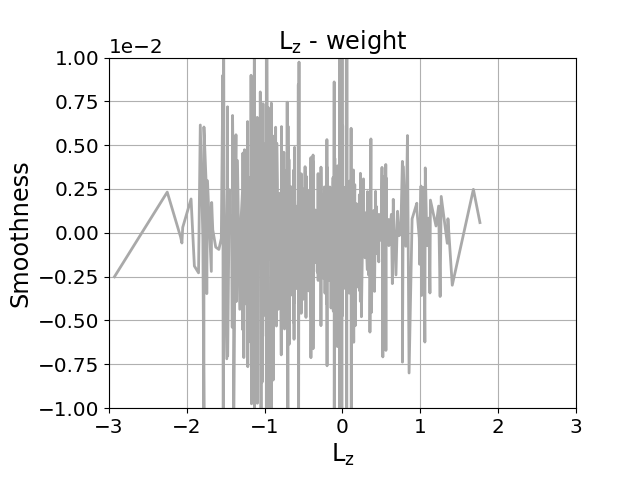}  & \includegraphics[width=50mm]{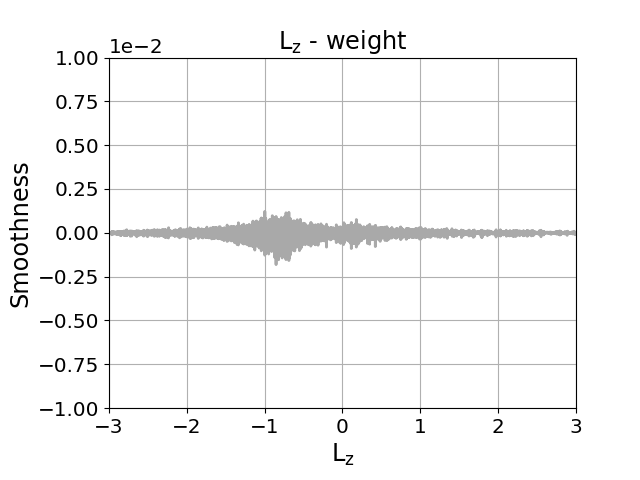}   & \includegraphics[width=50mm]{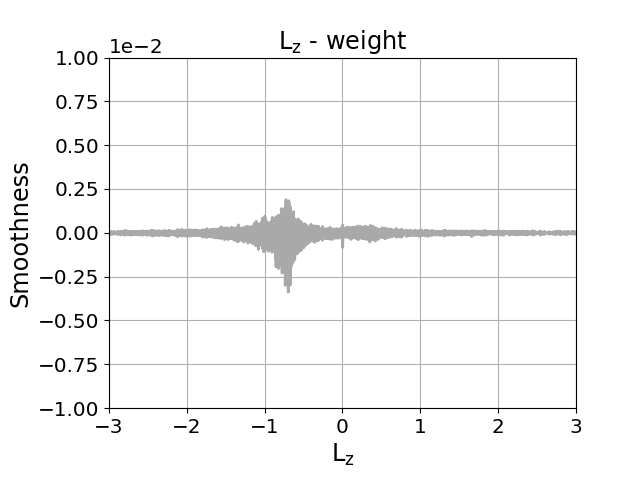}\\
	\end{tabular}
	
\medskip
Heavy weights penalization illustrated with NGC 4452 (Section \ref{sec:regSCHW}).  Smoothness is calculated as the second derivative of the weights with respect to the two integrals ($E$ and $L_z$).  As may be seen from the figure, the unpenalized SCHW plots show considerably more variation than penalized SCHW. Penalized SCHW, using a parameter value of $1.6 \times 10^3$, shows a similar degree of smoothness to M2M. 
\end{figure}

In order to ensure a more balanced comparison between M2M and SCHW, a regularized form of SCHW is used in which heavy weights are penalized.  This has a comparable end effect to the exponential smoothing used in M2M, and helps to ensure that the orbit weights of SCHW and M2M in integral space are smoothed to a similar degree.  This is illustrated in Figure \ref{fig:heavypen4452} using models created with the \atlas\ data for NGC 4452.  The other \atlas\ galaxies and the IllustrisTNG galaxies behave similarly.   Smoothness in Figure \ref{fig:heavypen4452} is calculated as the second derivative of the weights with respect to energy, and to the z-component of angular momentum.  As may be seen from the figure, the unpenalized SCHW plots are considerably less smooth than the penalized versions, with penalized SCHW exhibiting a similar degree of smoothness to M2M.  In addition, the impact of heavy weights penalization is to bring the SCHW weight range closer to that of M2M, and also to increase the number of active orbits in SCHW.  For NGC 4452, the M2M weight range is $(7.09\times10^{-7}, 1.71\times10{-3})$ and the number of active orbits is $15963$.  For SCHW, the range is $(2.35\times10^{-6}, 5.02\times10^{-2})$ with $635$ active orbits without penalization, and $(3.03\times10^{-9}, 1.01\times10^{-3})$ and $13106$ active orbits with penalization.  These points provide the background to our choice of penalized SCHW for M2M versus SCHW comparisons.

\subsection{$\chi^2$ and Elapsed Time Comparisons between M2M and SCHW}
\label{sec:M2MSCHW3I}
The focus in this section is on how well observed data are reproduced, and comparing modeling elapsed times.  Orbit classifications are considered in the next section.  M2M and SCHW models using both 3I and MDJV initial conditions and both data sets are created.  Table \ref{tab:AtlasMSchi2} contains the results for the \atlas\ galaxy data, and Table \ref{tab:TNGMSchi2} for the IllustrisTNG galaxy data.
In the tables, all the mean observable $\chi^2$ values are $\approx 1$ or less indicating that the observed data have been well-reproduced by both M2M and SCHW. An example is shown in Appendix \ref{app:obsrepro}.  M2M weight convergence values are $>93\%$ implying that convergence to constant weights, as expected in \citet{Syer1996}, is close to being achieved.  For \atlas\ , the modeling elapsed times show that M2M performs faster than SCHW by a factor of $\approx 4$ for 8000 orbit models and by $\approx 5$ for 16000 orbit models.  For IllustrisTNG data, M2M outperforms SCHW by a factor of $\approx 3$. Any of these performance improvement factors would benefit large surveys where there are thousands of galaxy data sets to be processed.  The NNLS contributions to SCHW elapsed times show that, with increasing model size (numbers of orbits in this case), the performance difference will become more marked depending on the algorithms within the NNLS implementation.
\begin{table}
	\centering
	\caption{\atlas\ M2M and SCHW 3I and MDJV Comparisons - $\chi2$ and Elapsed Times}
	\label{tab:AtlasMSchi2}
	\begin{tabular}{ccc|ccc|ccc}
		\hline
		&&Model&\multicolumn{3}{|c|}{Mean $\chi^2$ values}&M2M Weight&Elapsed\\
		Galaxy&Orbits&Duration&sb&los v1&los v2&Convergence&Time\\
		&&(hmdtu)&&&&&(s)\\
		\hline
		\textbf{3I- M2M}&&&&&&&\\
		NGC 1248&8000&200&0.31&0.27&0.63&99\%&358\\
		NGC 3838&16000&500&1.07&0.18&0.57&96\%&1065\\
		NGC 4452&16000&200&0.76&0.42&0.69&96\%&865\\
		NGC 4551&8000&200&0.41&0.19&0.76&100\%&217\\
		\textbf{3I- SCHW}&&&&&&&\\
		NGC 1248&8000&200&0.07&0.11&0.28&&1407 &(228)\\
		NGC 3838&16000&500&0.19&0.09&0.18&&5024 &(1379)\\
		NGC 4452&16000&200&0.30&0.13&0.25&&4555 &(1524)\\
		NGC 4551&8000&200&0.06&0.07&0.23&&900 &(238)\\
		\hline
		\textbf{MDJV - M2M}&&&&&&&\\
		NGC 1248&8000&200&1.06&0.11&0.30&96\%&561\\
		NGC 3838&16000&500&0.91&0.35&0.33&93\%&1229\\
		NGC 4452&16000&200&1.08&0.52&1.12&{94\%}&1008\\
		NGC 4551&8000&200&0.52&0.16&1.33&98\%&252\\
		\textbf{MDJV - SCHW}&&&&&&&\\
		NGC 1248&8000&200&0.05&0.12&0.36&&1486 &(225)\\
		NGC 3838&16000&500&0.12&0.13&0.21&&5422 &(1392)\\
		NGC 4452&16000&200&0.56&0.23&0.44&&4763 &(1422)\\
		NGC 4551&8000&200&0.10&0.07&0.32&&1012 &(267)\\
		\hline
	\end{tabular}
	
	\medskip
	$\chi2$ and elapsed time M2M and SCHW comparison using \atlas\ data (Section \ref{sec:M2MSCHW3I}).  All the mean observable $\chi^2$ values are $\approx 1$ or less indicating that the observed data have been well-reproduced by both M2M and SCHW. The M2M weight convergence is $>90\%$. The SCHW $\chi^2$ values are less than those for M2M due to over-fitting of the observed data (as expected).  M2M performs faster than SCHW in elapsed times by a factor of $\approx 4$ for 8000 orbit models and by $\approx 5$ for 16000 orbit models. The numbers in brackets are the NNLS contributions to the elapsed times. With increasing numbers of orbits, the performance difference will become more marked depending on the NNLS implementation used by SCHW.
\end{table}

\begin{table}
	\centering
	\caption{IllustrisTNG M2M and SCHW 3I and MDJV Comparisons - $\chi2$ and Elapsed Times}
	\label{tab:TNGMSchi2}
	\begin{tabular}{ccc|ccccc|ccc}
		\hline
		&&Model&\multicolumn{5}{|c|}{Mean $\chi^2$ values}&M2M Weight&Elapsed\\
		Galaxy&Orbits&Duration&sb&$h_1$&$h_2$&$h_3$&$h_4$&Convergence&Time\\
		&&(hmdtu)&&&&&&&(s)\\
		\hline
		\textbf{3I - M2M}&&&&&&&&&\\
		A0490&8000&200&0.94&0.72&1.10&0.66&1.23&96\%&1203\\
		A1090&8000&100&0.87&0.93&1.04&0.63&0.61&93\%&632\\
		A1190&8000&100&0.75&0.79&0.92&0.84&0.56&97\%&738\\
		A1290&8000&100&0.71&1.01&0.94&0.77&0.73&93\%&550\\
		A1390&8000&100&0.90&0.85&1.24&0.61&0.56&96\%&628\\
		\textbf{3I - SCHW}&&&&&&&&&\\
		A0490&8000&200&1.30&0.39&0.72&0.43&0.89&&3746 &(196)\\
		A1090&8000&100&0.54&0.49&0.59&0.25&0.31&&1948 &(200)\\
		A1190&8000&100&0.53&0.35&0.73&0.20&0.39&&2240 &(197)\\
		A1290&8000&100&0.33&0.37&0.41&0.23&0.38&&1816 &(200)\\
		A1390&8000&100&0.69&0.42&0.88&0.28&0.35&&2037 &(200)\\
		\hline
		\textbf{MDJV - M2M}&&&&&&&&&\\
		A0490&8000&100&1.04&0.75&0.79&0.54&1.07&96\%&753\\
		A1090&8000&100&0.78&0.88&1.01&0.52&0.43&95\%&776\\
		A1190&8000&100&0.61&0.60&0.58&0.43&0.53&96\%&858\\
		A1290&8000&100&1.21&0.70&0.68&0.64&0.54&97\%&665\\
		A1390&8000&100&0.85&0.81&1.06&0.46&0.44&96\%&776\\
		\textbf{MDJV - SCHW}&&&&&&&&&\\
		A0490&8000&100&0.43&0.20&0.21&0.21&0.42&&1939 &(235)\\
		A1090&8000&100&0.54&0.31&0.42&0.24&0.20&&2180 &(257)\\
		A1190&8000&100&0.26&0.22&0.31&0.15&0.21&&2336 &(267)\\
		A1290&8000&100&0.37&0.29&0.24&0.15&0.21&&1865 &(258)\\
		A1390&8000&100&0.48&0.33&0.44&0.20&0.25&&2193 &(258)\\
		\hline
	\end{tabular}
	
	\medskip
	$\chi2$ and elapsed time M2M and SCHW comparison using IllustrisTNG data (Section \ref{sec:M2MSCHW3I}).  All the observable $\chi^2$ values are $\approx 1$ or less indicating that the observed data have been well-reproduced by both M2M and SCHW with, for M2M, the weight convergence being $>90\%$.  M2M outperforms SCHW by a factor of $\approx 3$ in elapsed times.  As in Table \ref{tab:AtlasMSchi2}, the numbers in brackets are the NNLS contributions to the elapsed times.
\end{table}

\subsection{Orbit Classification Comparisons between M2M and SCHW}
\label{sec:orbitM2MSCHW}
The comparisons between orbit classification results from modeling with M2M and SCHW are shown in Table \ref{tab:3IMDJVcomp}. M2M and SCHW are producing similar orbit classifications from the same observed data and initial conditions, with \atlas\ and MDJV showing the most variation in classification and IllustrisTNG and MDJV showing the least.  For the same initial conditions, it is to be expected that both M2M and SCHW will yield similar orbit classifications as they have essentially the same log likelihood to maximize to reproduce the observed data.
\begin{table}
	\centering
	\caption{M2M versus SCHW Comparisons for 3I and MDJV - Orbit Classifications}
	\label{tab:3IMDJVcomp}	
	
	\begin{tabular}{c|ccccc|ccccc}
		\hline
		&\multicolumn{5}{|c|}{3I Initial Conditions}&\multicolumn{5}{|c}{MDJV Initial Conditions}\\
		Galaxy&Active&\multicolumn{4}{c|}{Orbit Classification}&Active&\multicolumn{4}{c}{Orbit Classification}\\
		&Orbits&Hot&Warm&Cold&C-R&Orbits&Hot&Warm&Cold&C-R\\
		\hline
		\textbf{Atlas3D – M2M}&&&&&&&&&&\\
		NGC 1248&7977  &{0.18}&0.41&0.23&\rjlb{0.17}                &8000  &{0.24}&0.43&\rjlb{0.21}&{0.12}\\
		NGC 3838&14098 &\rjlb{0.16}&{0.37}&\rjlb{0.36}&\rjlb{0.11}  &13742 &\rjlb{0.27}&\rjlb{0.51}&{0.16}&{0.06}\\
		NGC 4452&15896 &{0.17}&{0.31}&\rjlb{0.34}&{0.17}  &15389 &\rjlb{0.25}&{0.49}&{0.12}&\rjlb{0.14}\\
		NGC 4551&7889  &\rjlb{0.26}&{0.39}&{0.15}&{0.19}  &7882  &\rjlb{0.39}&{0.46}&{0.03}&\rjlb{0.13}\\
		\textbf{Atlas3D – SCHW}&&&&&&&&&&\\
		NGC 1248&7878  &{0.19}&{0.42}&{0.25}&\rjlb{0.13}  &7804  &{0.25}&{0.45}&\rjlb{0.17}&{0.13}\\
		NGC 3838&10484 &\rjlb{0.22}       &{0.39}&\rjlb{0.31}&\rjlb{0.07}  &12440 &\rjlb{0.23}       &\rjlb{0.56}&{0.15}&{0.07}\\
		NGC 4452&13106 &0.19       &{0.31}&\rjlb{0.31}&{0.18}  &8678  &\rjlb{0.19}       &{0.51}&{0.13}&\rjlb{0.17}\\
		NGC 4551&7771  &\rjlb{0.30}&{0.38}&{0.13}&{0.18}  &7225  &\rjlb{0.33}&{0.48}&{0.03}&\rjlb{0.16}\\
		\hline
		\textbf{IllustrisTNG – M2M}&&&&&&&&&&\\
		A0490&8000 &0.36       &\rjlb{0.28}&{0.13}&\rjlb{0.23}  &8000 &0.36       &{0.34}&{0.09}&{0.21}\\
		A1090&8000 &{0.30}&{0.43}&\rjlb{0.15}&{0.12}  &8000 &{0.34}&{0.46}&{0.09}&{0.11}\\
		A1190&8000 &{0.39}&\rjlb{0.40}       &{0.08}&{0.13}  &8000 &{0.42}&0.41       &{0.04}&{0.13}\\
		A1290&8000 &{0.31}&{0.34}&{0.20}&{0.14}  &7977 &{0.34}&{0.37}&{0.15}&{0.14}\\
		A1390&8000 &0.33       &{0.42}&{0.12}&{0.13}  &8000 &0.34       &{0.46}&{0.06}&{0.13}\\
		\textbf{IllustrisTNG - SCHW}&&&&&&&&&&\\
		A0490&4529 &0.38       &\rjlb{0.31}&{0.14}&\rjlb{0.18}  &6084 &0.37       &{0.34}&{0.09}&{0.20}\\
		A1090&5315 &{0.30}&{0.41}&\rjlb{0.18}&{0.11}  &6906 &{0.35}&{0.46}&{0.10}&{0.10}\\
		A1190&5581 &{0.40}&\rjlb{0.37}&{0.10}&{0.13}  &7379 &{0.43}&{0.40}&{0.04}&{0.13}\\
		A1290&5135 &{0.30}&       0.36&{0.21}&{0.13}  &6984 &{0.34}&0.38       &{0.15}&{0.13}\\
		A1390&5739 &0.33       &{0.40}&{0.13}&{0.13}  &7219 &0.35       &{0.46}&{0.06}&{0.12}\\
		\hline
	\end{tabular}
	
	\medskip
	Orbit classification comparisons for M2M and SCHW with both 3I and MDJV initial conditions (see Section \ref{sec:orbitM2MSCHW}).  Comparisons in the table are made vertically, first between M2M and SCHW using \atlas\ data, and then using IllustrisTNG data.  Classification differences of $\geq 3\%$ between M2M and SCHW are indicated in \rjlb{blue}. M2M and SCHW are producing similar orbit classifications from the same observed data, with \atlas\ and MDJV showing the most variation (top right `partition') and IllustrisTNG and MDJV showing the least (bottom right `partition').  The largest individual variation is $6\%$ and occurs in the hot classifications for \atlas\ data.
\end{table}

Orbit classifications are compiled from individual orbits and, given that here the same sets of orbits (initial conditions) are used for M2M and SCHW modeling, whether or not M2M and SCHW are using the same orbits in the same classifications is investigated.  Table \ref{tab:M2MSCHWorbitscomp} shows how many orbits are common to both modeling methods, and how many are specific to one of M2M or SCHW. Comparatively few are specific to SCHW only, with substantially more being specific to M2M.  This is readily explainable: orbits which are zero weighted by SCHW appear in the M2M only figures.
\begin{table}
	\centering
	\caption{M2M versus SCHW Orbit Comparison}
	\label{tab:M2MSCHWorbitscomp}	
	
	\begin{tabular}{c|ccccc|ccccc}
		\hline
		&\multicolumn{5}{|c|}{3I Initial Conditions}&\multicolumn{5}{|c}{MDJV Initial Conditions}\\
		Galaxy&\multicolumn{2}{|c}{Active Orbits}&Common&M2M&SCHW&\multicolumn{2}{|c}{Active Orbits}&Common&M2M&SCHW\\
		&M2M&SCHW&Orbits&only&only&M2M&SCHW&Orbits&only&only\\
		\hline
		\textbf{\atlas}&&&&&&&&&&\\
		NGC 1248&7992&7886&7843&149&3&7985&7812&7759&226&9\\
		NGC 3838&15947&10528&10087&4992&290&15763&12475&12241&2645&113\\
		NGC 4452&15963&13106&12910&3053&20&15992&8678&8587&7251&7\\
		NGC 4551&7975&7780&7726&232&22&7955&7225&7158&797&7\\
		\hline
		\textbf{IllustrisTNG}&&&&&&&&&&\\
		A0490&7993&4563&4519&3474&0&8000&6087&6032&1968&0\\
		A1090&8000&5315&5259&2741&0&8000&6906&6855&1145&0\\
		A1190&8000&5581&5523&2416&0&8000&7390&7339&661&0\\
		A1290&8000&5135&5074&2926&0&8000&6992&6926&1045&11\\
		A1390&8000&5739&5684&2316&0&8000&7219&7165&823&0\\
		\hline
	\end{tabular}
	
	\medskip
M2M versus SCHW orbit comparisons (Section \ref{sec:orbitM2MSCHW}) using both sets of initial conditions and both data sets.  The comparison for every galaxy's data shows the number of orbits which are common to both M2M and SCHW models, and those that are particular to just one model.  Zero-weighted orbits in SCHW models, if they are used by M2M, are accounted for in the M2M only subset.  This explains the comparatively small values in the SCHW only column, and the much larger values in the M2M only column.  The common orbits are examined in more detail in Figure \ref{fig:commonorbit}.
\end{table}

\begin{figure}[h]
	\centering
	\caption{M2M and SCHW Common Orbits Weight Comparison for \atlas\ Data}
	\label{fig:commonorbit}
	\begin{tabular}{cM{45mm}M{45mm}M{45mm}}
		 & M2M vs SCHW Weights & Bland-Altman & Log Weights Ratio\\
		NGC 1248 & \includegraphics[width=45mm]{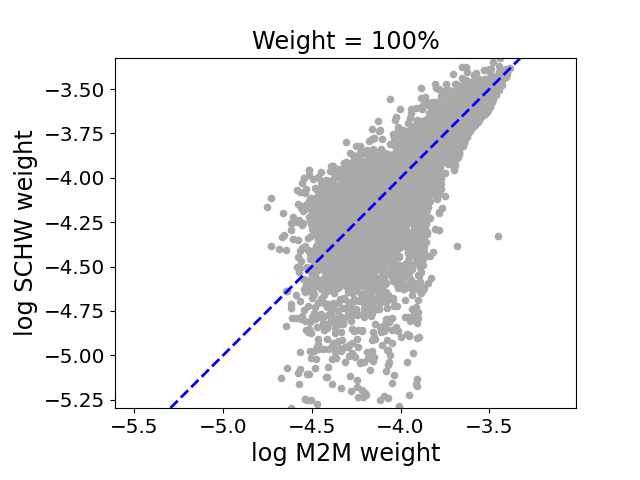}  & \includegraphics[width=45mm]{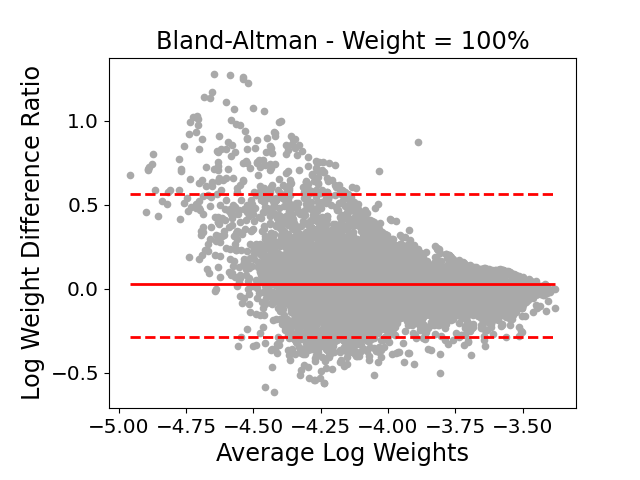}  & \includegraphics[width=45mm]{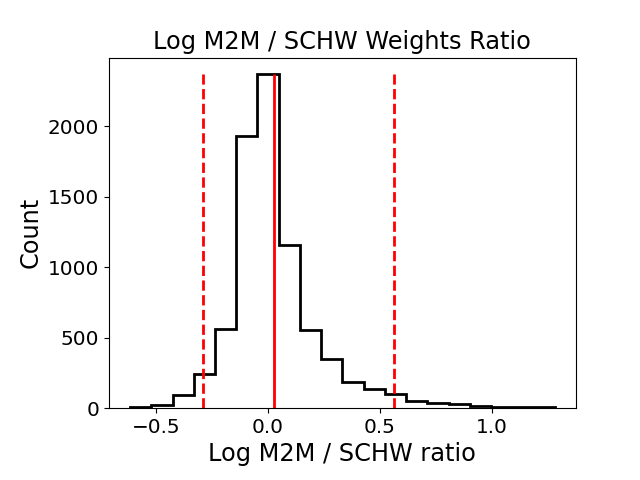}\\
		NGC 3838 & \includegraphics[width=45mm]{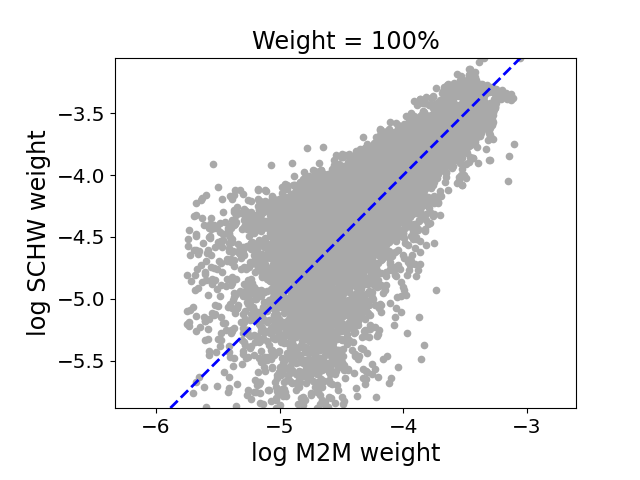}  & \includegraphics[width=45mm]{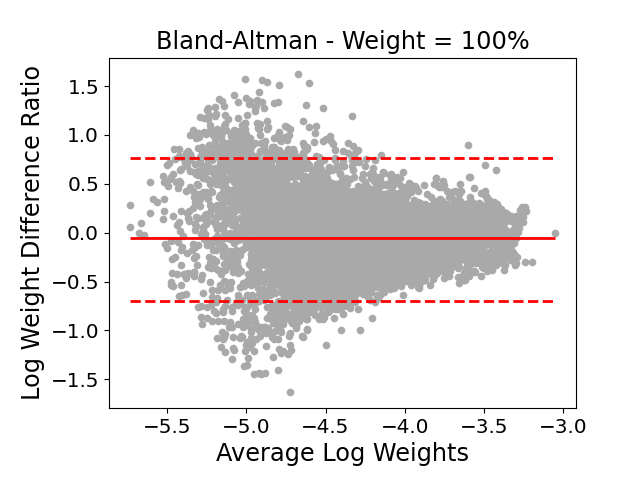}  & \includegraphics[width=45mm]{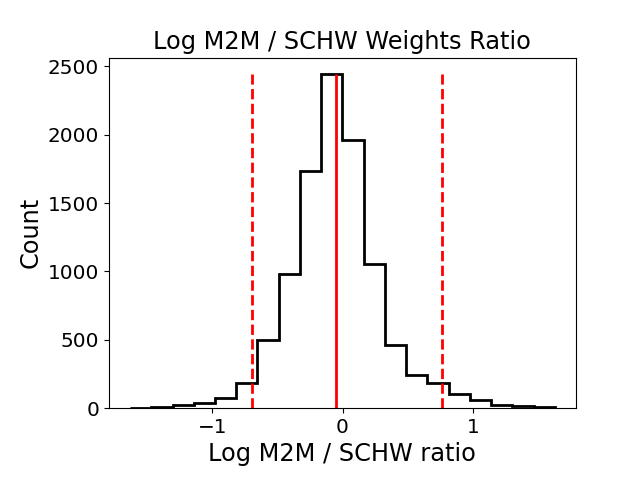}\\
		NGC 4452 & \includegraphics[width=45mm]{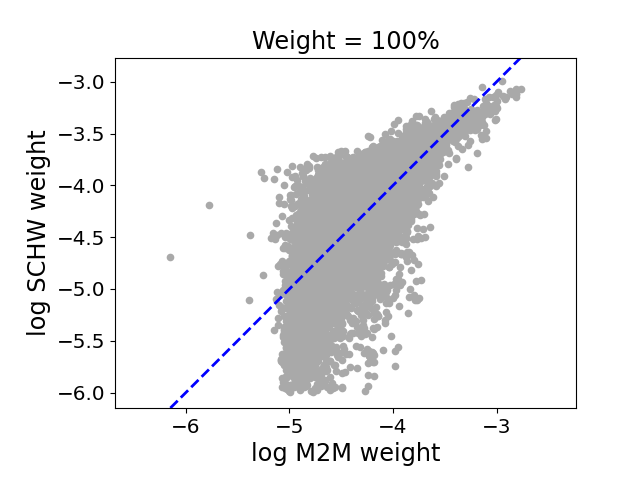}  & \includegraphics[width=45mm]{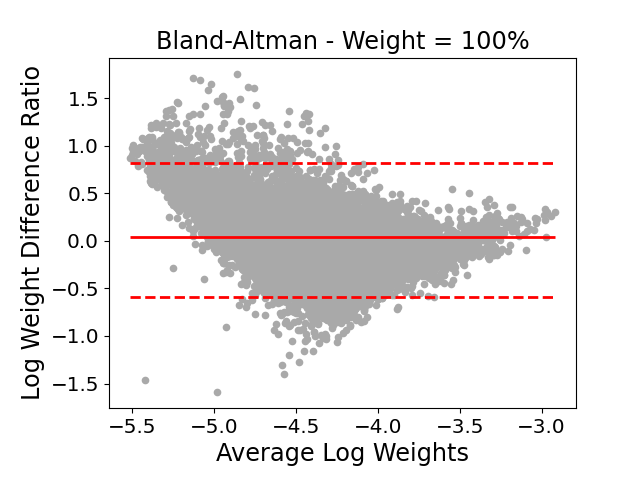}  & \includegraphics[width=45mm]{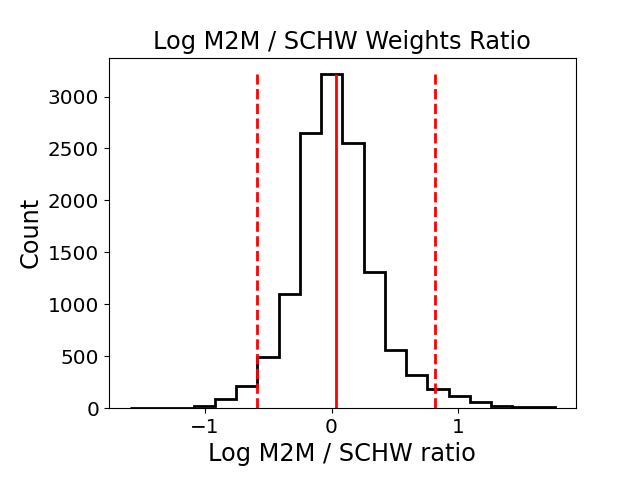}\\
		NGC 4551 & \includegraphics[width=45mm]{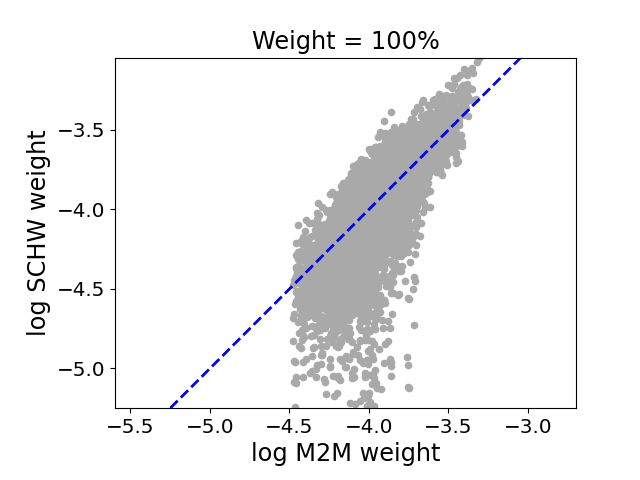}  & \includegraphics[width=45mm]{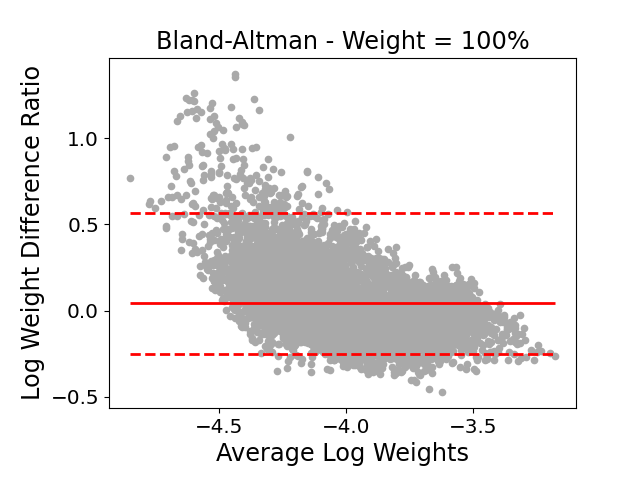}  & \includegraphics[width=45mm]{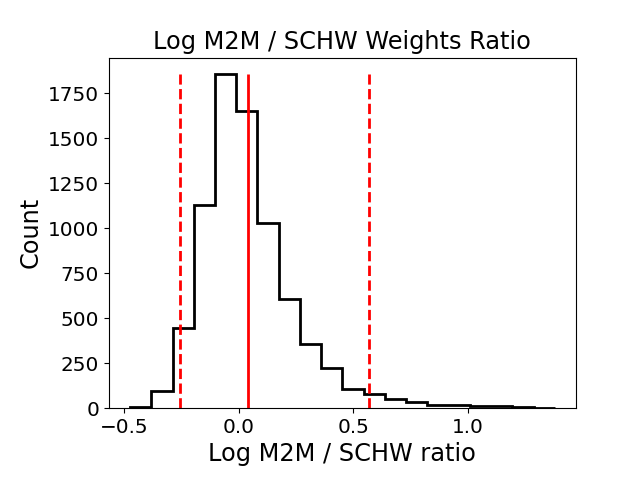}\\
	\end{tabular}
	
	\medskip
M2M and SCHW common orbits weight comparisons for \atlas\ data (Section \ref{sec:orbitM2MSCHW}). 3I initial conditions are used for the models.  The left hand column shows scatter plots of the weights of the orbits, with the \rjlb{blue} dashed line indicating equality of the orbit weights.  The key point to note is that the weights are not tightly positioned close to the equality line: even though the orbits used are the same, M2M and SCHW are in general not allocating the same weights to the orbits.  The Bland-Altman plots \citep{BA1983} of the average log weight for an orbit against the log difference between its M2M weight and its SCHW weight show that the mean log difference (\rjlr{red} solid line) is approximately zero indicating that there is no significant bias towards either method.  In addition, weight difference is larger at lower average weights than with larger average weights.  The \rjlr{red} dashed lines show the $2.5\%$ and $97.5\%$ percentiles with $95\%$ of the difference ratios lying between the lines. The right hand plots show the distributions of the weight ratios. The \rjlr{red} lines have the same meaning as in the Bland-Altman plots.
\end{figure}

The next step is to look in detail at the common orbits, to investigate whether individual orbits are weighted similarly by M2M and SCHW.
Figure \ref{fig:commonorbit} contains various plots from considering just the \atlas\ data set modeled with 3I initial conditions.  
The left hand scatter plots show the paired orbit weights arising from M2M and SCHW modeling, with the \rjlb{blue} dashed line indicating equality of the orbit weights.  The key point to note is that the weights are not tightly positioned close to the equality line: even though the orbits used are the same, M2M and SCHW are in general not allocating the same weights to the orbits.  
The central Bland-Altman plots of the average log weight for an orbit against the log difference between its M2M weight and its SCHW weight show that the mean log difference (\rjlr{red} solid line) is approximately zero (absolute values $<10^{-2}$) indicating that there is no significant bias towards either method.  In addition, weight differences  are larger at lower average weights than with higher average weights.  The \rjlr{red} dashed lines show the $2.5\%$ and $97.5\%$ percentiles with $95\%$ of the difference ratios lying between the lines. Expressing the values of the percentiles as percentages, they are equivalent to saying that the weight differences range from a few tens of percent lower to a few hundred percent higher.  
The right hand plots show the distributions of the weight ratios. The \rjlr{red} lines have the same meaning as in the Bland-Altman plots.

Finally, it is useful to note that if in a model orbit weights are sorted into descending order, then just how many orbits contribute to the heaviest $25\%$, $50\%$, $75\%$ and $100\%$ by weight can be established.  Using 3I initial conditions, the approximate profile for \atlas\ is [$8\%$, $22\%$, $44\%$, $100\%$] of the orbits, and for IllustrisTNG [$4\%$, $14\%$, $35\%$, $100\%$]. What is then clear is the first $25\%$ of the total weight involves a small number of heavily weighted orbits, while over $50\%$ orbits (lightly weighted) make up the final $25\%$.
Both M2M and SCHW are affected in the same way.

\subsection{Comparisons between 3I and MDJV}
\label{sec:orbit3IMDJV}
The orbit classification results achieved with 3I and MDJV initial conditions are compared for M2M and SCHW using both 3I and MDJV initial conditions.  The comparisons are shown in Table \ref{tab:M2MSCHW3IMDJVcomp}.  The numerical values in Table \ref{tab:M2MSCHW3IMDJVcomp} are the same as Table \ref{tab:3IMDJVcomp} but this time the comparisons are made horizontally (3I versus MDJV) rather than vertically (M2M versus SCHW).  From the coloring used, it should be apparent from Table \ref{tab:M2MSCHW3IMDJVcomp} that 3I and MDJV initial conditions do not produce the same orbit classifications.  The only class where there is some agreement is with the counter-rotating orbits.

\begin{table}
	\centering
	\caption{3I versus MDJV Comparisons for M2M and SCHW - Orbit Classifications}
	\label{tab:M2MSCHW3IMDJVcomp}	
	
	\begin{tabular}{c|ccccc|ccccc}
		\hline
		&\multicolumn{5}{|c|}{3I Initial Conditions}&\multicolumn{5}{|c}{MDJV Initial Conditions}\\
		Galaxy&Active&\multicolumn{4}{c|}{Orbit Classification}&Active&\multicolumn{4}{c}{Orbit Classification}\\
		&Orbits&Hot&Warm&Cold&C-R&Orbits&Hot&Warm&Cold&C-R\\
		\hline
		\textbf{Atlas3D – M2M}&&&&&&&&&&\\
		NGC 1248&7977  &\rjlr{0.18}&0.41&0.23&\rjlr{0.17}                &8000  &\rjlr{0.24}&0.43&0.21&\rjlr{0.12}\\
		NGC 3838&14098 &\rjlr{0.16}&\rjlr{0.37}&\rjlr{0.36}&\rjlr{0.11}  &13742 &\rjlr{0.27}&\rjlr{0.51}&\rjlr{0.16}&\rjlr{0.06}\\
		NGC 4452&15896 &\rjlr{0.17}&\rjlr{0.31}&\rjlr{0.34}&\rjlr{0.17}  &15389 &\rjlr{0.25}&\rjlr{0.49}&\rjlr{0.12}&\rjlr{0.14}\\
		NGC 4551&7889  &\rjlr{0.26}&\rjlr{0.39}&\rjlr{0.15}&\rjlr{0.19}  &7882  &\rjlr{0.39}&\rjlr{0.46}&\rjlr{0.03}&\rjlr{0.13}\\
		\textbf{Atlas3D – SCHW}&&&&&&&&&&\\
		NGC 1248&7878  &\rjlr{0.19}&\rjlr{0.42}&\rjlr{0.25}&{0.13}  &7804  &\rjlr{0.25}&\rjlr{0.45}&\rjlr{0.17}&{0.13}\\
		NGC 3838&10484 &0.22       &\rjlr{0.39}&\rjlr{0.31}&{0.07}  &12440 &0.23       &\rjlr{0.56}&\rjlr{0.15}&{0.07}\\
		NGC 4452&13106 &0.19       &\rjlr{0.31}&\rjlr{0.31}&{0.18}  &8678  &0.19       &\rjlr{0.51}&\rjlr{0.13}&{0.17}\\
		NGC 4551&7771  &\rjlr{0.30}&\rjlr{0.38}&\rjlr{0.13}&{0.18}  &7225  &\rjlr{0.33}&\rjlr{0.48}&\rjlr{0.03}&{0.16}\\
		\hline
		\textbf{IllustrisTNG – M2M}&&&&&&&&&&\\
		A0490&8000 &0.36       &\rjlr{0.28}&\rjlr{0.13}&{0.23}  &8000 &0.36       &\rjlr{0.34}&\rjlr{0.09}&{0.21}\\
		A1090&8000 &\rjlr{0.30}&\rjlr{0.43}&\rjlr{0.15}&{0.12}  &8000 &\rjlr{0.34}&\rjlr{0.46}&\rjlr{0.09}&{0.11}\\
		A1190&8000 &\rjlr{0.39}&0.40       &\rjlr{0.08}&{0.13}  &8000 &\rjlr{0.42}&0.41       &\rjlr{0.04}&{0.13}\\
		A1290&8000 &\rjlr{0.31}&\rjlr{0.34}&\rjlr{0.20}&{0.14}  &7977 &\rjlr{0.34}&\rjlr{0.37}&\rjlr{0.15}&{0.14}\\
		A1390&8000 &0.33       &\rjlr{0.42}&\rjlr{0.12}&{0.13}  &8000 &0.34       &\rjlr{0.46}&\rjlr{0.06}&{0.13}\\
		\textbf{IllustrisTNG - SCHW}&&&&&&&&&&\\
		A0490&4529 &0.38       &\rjlr{0.31}&\rjlr{0.14}&{0.18}  &6084 &0.37       &\rjlr{0.34}&\rjlr{0.09}&{0.20}\\
		A1090&5315 &\rjlr{0.30}&\rjlr{0.41}&\rjlr{0.18}&{0.11}  &6906 &\rjlr{0.35}&\rjlr{0.46}&\rjlr{0.10}&{0.10}\\
		A1190&5581 &\rjlr{0.40}&\rjlr{0.37}&\rjlr{0.10}&{0.13}  &7379 &\rjlr{0.43}&\rjlr{0.40}&\rjlr{0.04}&{0.13}\\
		A1290&5135 &\rjlr{0.30}&       0.36&\rjlr{0.21}&{0.13}  &6984 &\rjlr{0.34}&0.38       &\rjlr{0.15}&{0.13}\\
		A1390&5739 &0.33       &\rjlr{0.40}&\rjlr{0.13}&{0.13}  &7219 &0.35       &\rjlr{0.46}&\rjlr{0.06}&{0.12}\\
		\hline
	\end{tabular}
	
\medskip
The orbit classification comparisons for 3I and MDJV initial conditions (Section \ref{sec:orbit3IMDJV}).  Comparisons in the table are made horizontally with classification differences of $\geq 3\%$ between 3I and MDJV being indicated in \rjlr{red}. 3I and MDJV do not produce the same orbit classifications, though it is interesting that there is less disagreement on the counter-rotating orbit classification.
\end{table}

\subsection{Comparison with \citet{Xu2019}}
\label{sec:Xucomp}
The orbit classification comparisons for M2M and SCHW using IllustrisTNG data in Section \ref{sec:orbitM2MSCHW} are compared with \citet{Xu2019}, and the results recorded in Table \ref{tab:Xucomp}.  Apart from the A0490 M2M model, all the other models have differences with \citet{Xu2019} in the hot, warm and cold classifications and this is considered further in the Discussion (Section \ref{sec:discussion}).  Apart from the counter-rotating classification for the A0490 SCHW model, all other counter-rotating classifications agree, to within $3\%$, with \citet{Xu2019}. The differences in the MDJV cold column reflect the fact that MDJV produces fewer cold orbits than 3I, and does not meet the orbital needs of the observed data.
\begin{table}
	\centering
	\caption{IllustrisTNG M2M and SCHW Comparisons with \citet{Xu2019} - Orbit Classifications}
	\label{tab:Xucomp}	
	
	\begin{tabular}{c|cccc|cccc|cccc}
		\hline
		&\multicolumn{4}{|c|}{3I Initial Conditions}&\multicolumn{4}{|c}{\citet{Xu2019}}&\multicolumn{4}{|c}{MDJV Initial Conditions}\\
		Galaxy&\multicolumn{4}{c|}{Orbit Classification}&\multicolumn{4}{c|}{Orbit Classification}&\multicolumn{4}{c}{Orbit Classification}\\
		&Hot&Warm&Cold&C-R&Hot&Warm&Cold&C-R&Hot&Warm&Cold&C-R\\
		\hline
		\textbf{IllustrisTNG – M2M}&&&&&&&&\\
		A0490 &0.36       &{0.28}&{0.13}&{0.23}   &0.35       &{0.28}&{0.14}&{0.23} &0.36       &\rjlb{0.34}&\rjlb{0.09}&{0.21}\\
		A1090 &{0.30}&\rjlr{0.43}&{0.15}&{0.12}        &{0.28}&{0.46}&{0.15}&{0.10} &\rjlb{0.34}&{0.46}&\rjlb{0.09}&{0.11}\\
		A1190 &\rjlr{0.39}&\rjlr{0.40}  &\rjlr{0.08}&{0.13} &{0.33}&{0.43}       &{0.11}&{0.12} &\rjlb{0.42}&0.41       &\rjlb{0.04}&{0.13}\\
		A1290 &{0.31}&\rjlr{0.34}&{0.20}&{0.14}        &{0.32}&{0.37}&{0.18}&{0.13} &{0.34}&{0.37}&\rjlb{0.15}&{0.14}\\
		A1390 &\rjlr{0.33}       &\rjlr{0.42}&\rjlr{0.12}&{0.13}   &{0.29}       &{0.45}&{0.15}&{0.12} &\rjlb{0.34}       &{0.46}&\rjlb{0.06}&{0.13}\\
		\textbf{IllustrisTNG - SCHW}&&&&&&&&\\
		A0490&\rjlr{0.38}       &\rjlr{0.31}&{0.14}&\rjlr{0.18}   &{0.35}       &{0.28}&{0.14}&{0.23} &0.37       &\rjlb{0.34}&\rjlb{0.09}&\rjlb{0.20}\\
		A1090&{0.30}&\rjlr{0.41}&\rjlr{0.18}&{0.11}        &{0.28}&{0.46}&{0.15}&{0.10}&\rjlb{0.35}&{0.46}&\rjlb{0.10}&{0.10}\\
		A1190&\rjlr{0.40}&\rjlr{0.37}&{0.10}&{0.13}        &{0.33}&{0.43}&{0.11}&{0.12} &\rjlb{0.43}&\rjlb{0.40}&\rjlb{0.04}&{0.13}\\
		A1290&{0.30}&       0.36&\rjlr{0.21}&{0.13}   &{0.32}&0.37       &{0.18}&{0.13} &{0.34}&0.38       &\rjlb{0.15}&{0.13}\\
		A1390&\rjlr{0.33}       &\rjlr{0.40}&{0.13}&{0.13}   &{0.29}       &{0.45}&{0.15}&{0.12} &\rjlb{0.35}       &{0.46}&\rjlb{0.06}&{0.12}\\
		\hline
	\end{tabular}
	
	\medskip
	Orbit classification comparisons for M2M and SCHW with \citet{Xu2019} (see Section \ref{sec:Xucomp}).  Comparisons in the table are made horizontally between models and \citet{Xu2019}.  Classification differences of $\geq 3\%$ with \cite{Xu2019} are indicated in \rjlr{red} and \rjlb{blue}. Apart from the A0490 M2M model, all the other models have some differences with \citet{Xu2019}.  The values in the counter-rotating columns are notable for their lack of differences.  The MDJV cold column reflects differences between the initial conditions where MDJV has fewer cold orbits than 3I.
\end{table}

\subsection{Impacts of Changing Data Layout}
\label{sec:layout}
\begin{figure}[h]
	\centering
	\caption{Data Positions for NGC 4452}
	\label{fig:dataposn4452}
	\begin{tabular}{cc}
		\includegraphics[width=70mm]{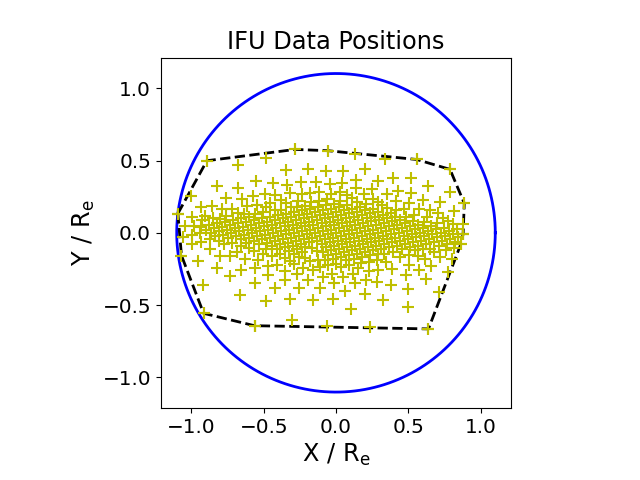}  & \includegraphics[width=70mm]{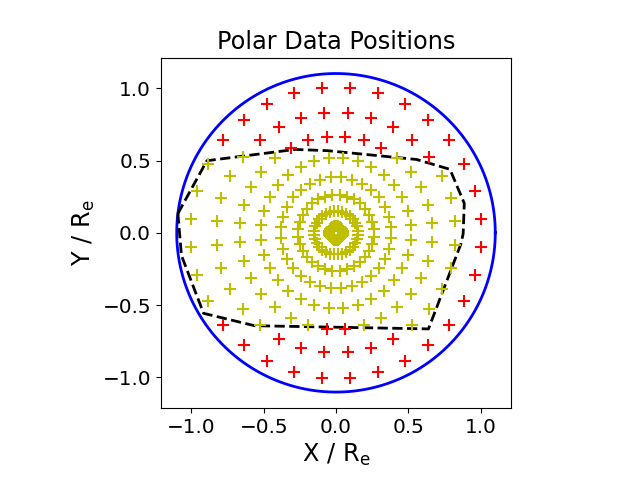} \\
	\end{tabular}
	
	\medskip
IFU and polar data positions for NGC 4452 kinematical data (Section \ref{sec:layout}).  The left hand plot shows the IFU data positions in yellow with the black dashed line showing the convex hull enclosing the data points.  The blue circle has the minimum radius necessary to surround all data points.  This radius gives the extent of the polar grid in the right hand plot.  In this plot, the yellow data points have data values interpolated from the IFU data values.  The red points are the points in the polar grid that are outside of the convex and have no data values.
\end{figure}

In this section, the impact of interpolating the IFU kinematical data from their Voronoi cells onto a polar grid is considered.  The polar grid chosen has 8 divisions logarithmically in radius and 32 in angle (making up $2\pi$ radians).  The Voronoi observed data are interpolated onto this $(8, 32)$ polar grid using `nearest point' interpolation (see Figure \ref{fig:dataposn4452}).  The modeling runs are limited to M2M with 3I initial conditions and \atlas\ galaxy data only: no use is made of SCHW, MDJV or IllustrisTNG data for this part of the investigation.

No issues with observable $\chi^2$ values were noted and the M2M weights converged as required (Table \ref{tab:VPchi2}).  
\begin{table}
	\centering
	\caption{Voronoi and Polar Kinematical Data - $\chi2$ Values}
	\label{tab:VPchi2}
	\begin{tabular}{cc|ccc|c}
		\hline
		&&&Mean $\chi^2$ values&&M2M Weight\\
		Galaxy&Orbits&sb&los v1&los v2&Convergence\\
		\hline
		\textbf{Polar kinematical data}&&&&&\\
		NGC 1248&8000&0.38&0.23&0.25&99\%\\
		NGC 3838&16000&1.01&0.51&0.50&92\%\\
		NGC 4452&16000&1.16&0.31&1.07&96\%\\
		NGC 4551&8000&0.87&0.24&1.17&98\%\\
		\textbf{Voronoi kinematical data}&&&&&\\
		NGC 1248&8000&0.31&0.27&0.63&99\%\\
		NGC 3838&16000&1.07&0.18&0.57&96\%\\
		NGC 4452&16000&0.76&0.42&0.69&96\%\\
		NGC 4551&8000&0.41&0.19&0.76&100\%\\
		\hline
	\end{tabular}
	
	\medskip
	The observable mean $\chi^2$ values from the M2M \atlas\ polar versus Voronoi cells comparison (Section \ref{sec:layout}).  All the $\chi^2$ values are $\approx 1$ or less, and the weight convergence values are $>90\%$ indicting that M2M has reproduced the observed data as expected.  
\end{table}
Orbit classification values are in Table \ref{tab:VPcirc}. Nine differences $\geq 3\%$ are recorded, with the maximum difference being $9\%$.
Overall, it appears that the geometric arrangement of the observed data does influence the orbit classifications obtained.  Given the results obtained in the investigation as a whole, it is considered unlikely that the geometric effect is M2M specific and does not affect SCHW as well.
\begin{table}
	\centering
	\caption{Voronoi and Polar Kinematics - Orbit Classification}
	\label{tab:VPcirc}	
	\begin{tabular}{ccc|cccc}
		\hline
		&&Active&\multicolumn{4}{|c}{Orbit Classification}\\
		Galaxy&Orbits&Orbits&Hot&Warm&Cold&C-R\\
		\hline
		\textbf{Polar kinematics}&&&&&&\\
		NGC 1248&8000&7782&\rjlr{0.21}&0.41&0.23&0.15\\
		NGC 3838&16000&12749&\rjlr{0.23}&\rjlr{0.40}&\rjlr{0.27}&0.11\\
		NGC 4452&16000&16000&0.18&\rjlr{0.28}&\rjlr{0.28}&\rjlr{0.25}\\
		NGC 4551&8000&7824&\rjlr{0.29}&0.41&\rjlr{0.12}&0.18\\
		\textbf{Voronoi kinematics}&&&&&&\\
		NGC 1248&8000&7977&\rjlr{0.18}&0.41&0.23&0.17\\
		NGC 3838&16000&14098&\rjlr{0.16}&\rjlr{0.37}&\rjlr{0.36}&0.11\\
		NGC 4452&16000&15896&0.17&\rjlr{0.31}&\rjlr{0.34}&\rjlr{0.17}\\
		NGC 4551&8000&7889&\rjlr{0.26}&0.39&\rjlr{0.15}&0.19\\
		\hline
	\end{tabular}
	
	\medskip
	The orbit classification values from the M2M \atlas\ polar versus Voronoi cells comparison (Section \ref{sec:layout}).  Table values in \rjlr{red} indicate a classification difference of $\geq 3\%$.  In total, there are nine \rjlr{red} values suggesting that orbit classifications are dependent on the geometric arrangement of the observed data.  The largest difference in classification is $9\%$. 
\end{table}

The impact of changing the surface brightness layout is now briefly considered.  Taking one galaxy's \atlas\ data (NGC 4551), 3I initial conditions, and modeling with M2M, the following polar grid layouts are employed: $(16, 8)$ and $(8, 16)$ with uniform radial intervals, and $(16, 16)$ with uniform and logarithmic radial intervals.  This gives four different surface brightness layouts.  The kinematical data are unmodified from the IFU data used for the NGC 4551 3I modeling run in Table \ref{tab:AtlasMSchi2}.  The results obtained for the four modeling runs are
\begin{enumerate}
	\item the mean $\chi^2$ values for surface brightness range from $0.14$ to $0.41$,
	\item the $\chi^2$ values for the first and second velocity moments are the same for all runs,
	\item M2M weight convergence is $>99\%$, and
	\item the orbit classifications are the same for all runs to within $1\%$.
\end{enumerate}
In short, varying the surface brightness polar layout in isolation has had little effect on the modeling results achieved.  This clearly differs from the kinematical data results described above.

\section{Discussion}
\label{sec:discussion}

The Results (Section \ref{sec:results}) contain many tables and much detail so the main findings are summarized below.
\begin{enumerate}
	\item M2M and SCHW consistently achieve mean observable $\chi^2 \approx 1$ or less for both data sets and initial conditions. Related, M2M weight convergence is $> 90\%$.  In other words, both methods are able to reproduce observed data to an acknowledged, acceptable level (see Tables \ref{tab:AtlasMSchi2} and \ref{tab:TNGMSchi2}).
	\item In terms of computer elapsed time, using the same computer configuration, M2M consistently outperforms SCHW.  M2M elapsed times are between three and five times faster than SCHW (see Tables \ref{tab:AtlasMSchi2} and \ref{tab:TNGMSchi2}).
	\item For both data sets, M2M and SCHW produce orbit classifications which are approximately the same for the same initial conditions (Table \ref{tab:3IMDJVcomp}), but are not the same for different initial conditions (Table \ref{tab:M2MSCHW3IMDJVcomp}).  The classifications are dependent on the initial conditions.
	\item Both M2M and SCHW using IllustrisTNG data as in \citet{Xu2019} do not produce orbit classifications which match those in \citet{Xu2019} (see Table \ref{tab:Xucomp}).  It therefore appears that the gravitational potential, observed data sets and initial conditions used by the methods are not able to produce good `orbit level' representations of the associated galaxies.  However, the classification values in \citet{Xu2019} are calculated from six dimensional phase space data, and are taken as the true classifications for the simulated galaxies.  The observed data for modeling taken from the simulated galaxies do not contain sufficient information to recreate these true classifications.  This is consistent with the data from real galaxies.  It is possible to use the \citet{Xu2019} classifications as model constraints and this was demonstrated in Long (2025).
	\item M2M and SCHW, using the same gravitational potentials and initial conditions, do not use exactly the same orbits in reproducing the same observed data.  For orbits which are common, these orbits do not have the same weights (see Figure \ref{fig:commonorbit}).
	\item Changing the data layout of surface brightness from one polar system to another does not appear to affect model results.  However, changing the kinematical data layout from IFU Voroni cells to a polar system does cause the orbit classification results to change (see Table \ref{tab:VPcirc}).
\end{enumerate}

Nothing has come out of the investigation which indicates that the performance benefit that comes from using M2M should not be taken advantage of.  The benefits to galaxy survey modeling are obvious.  In addition, M2M is not affected by the performance limitations of using NNLS.  Also, if the performance cost of NNLS were to be reduced to close to zero, M2M would still be faster.  All that is required is for researchers to continue to be comfortable with modeling schemes using low orbit numbers to reproduce galaxy data, and with creating models that do not necessarily truly represent the parent galaxy and may actually be non-physical (as another well-known scheme may do).  Do not forget that, regardless of regularization, orbit selection within the methods is purely numerical and not astrophysical.  

All comparisons between M2M and SCHW will, at some level, be implementation specific, but there seems to be little to be gained by seeking to address known modeling shortfalls in an SCHW-like context.  M2M is inherently more flexible and scalable than SCHW. Orbit weights are available within the model from the outset so more complex observables are easier to model or estimate.  Data uncertainties are an integral part of the M2M weight determination process, and are accommodated far more so than SCHW does in reaching the final solution.  Once designed, achieving a basic M2M implementation in approximately six to eight weeks can be accomplished by an experienced programmer using a high level language such as Python, and including use of parallelization.  Creating a GPU based implementation for M2M can be achieved using tooling such as PyTorch \citep{paszke2019}.  However to gain the full performance benefits, this will require careful assessment and design of the GPU parallelizable features of M2M.  Given that M2M employs a number of techniques that are used in machine learning, perhaps they might provide an alternative way forward and this will be investigated in the future by the authors

\section{Conclusions}
\label{sec:conclusions}
The objective set out in the Introduction to assess and compare the operational capabilities of M2M and SCHW has been met. Using the low orbit numbers typical of SCHW, both methods are able to replicate observed data to an acceptable level but M2M outperforms SCHW being several times faster.  This speed improvement makes M2M more desirable in a galaxy survey context than SCHW.  Both methods produce similar orbit classifications for a given set of initial conditions, but the classifications do appear to be particular to the initial conditions used.  In addition, the weights of the common orbits under-pinning the classifications differ and are method specific. Looking ahead, based on the results here and given its inherent flexibility, M2M is the more attractive option currently for stellar dynamical modeling in the 21st century, but could eventually be overtaken by machine learning based methods.  In the mean time, this paper acts as a flag to the stellar dynamical modeling community that M2M should be seriously considered as a way of alleviating some of the shortcomings of SCHW.

\begin{acknowledgements}
The author thanks Dandan Xu, Ling Zhu, Shude Mao and Yunpeng Jin for various fruitful discussions as the research recorded here progressed. 
\end{acknowledgements}

\bibliographystyle{raa}
\bibliography{ms2025-0459}

\appendix
\section{Example Observable Reproduction}
\label{app:obsrepro}
\begin{figure}[h]
	\centering
	\caption{M2M and SCHW Reproduction of Observables - NGC 4452}
	\label{fig:obsrepro4452}
	\begin{tabular}{M{45mm}M{45mm}M{45mm}}
		Observed Data & M2M & SCHW\\
		\includegraphics[width=45mm]{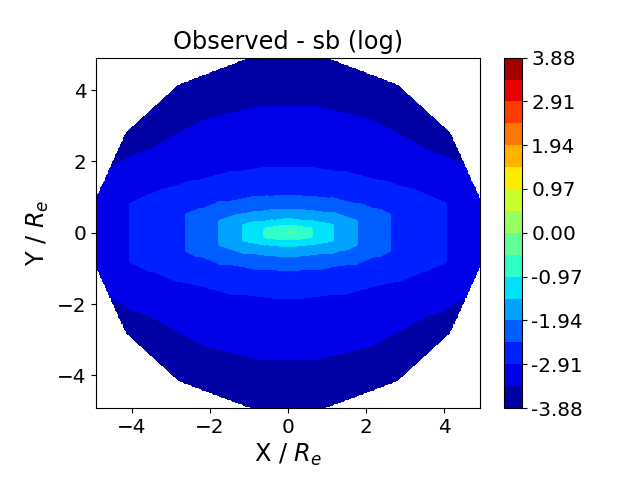}  & \includegraphics[width=45mm]{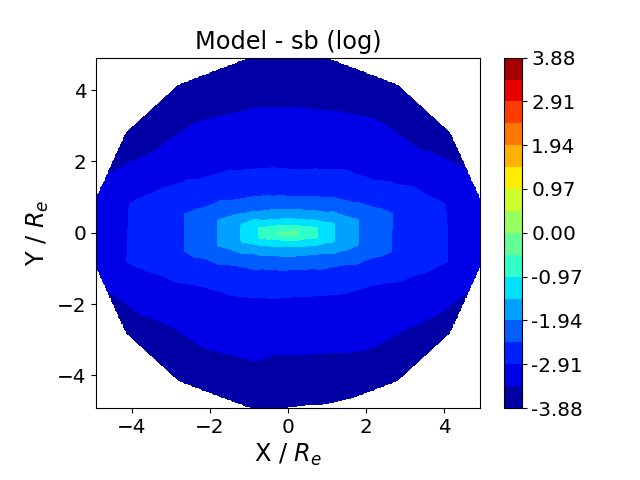}  & \includegraphics[width=45mm]{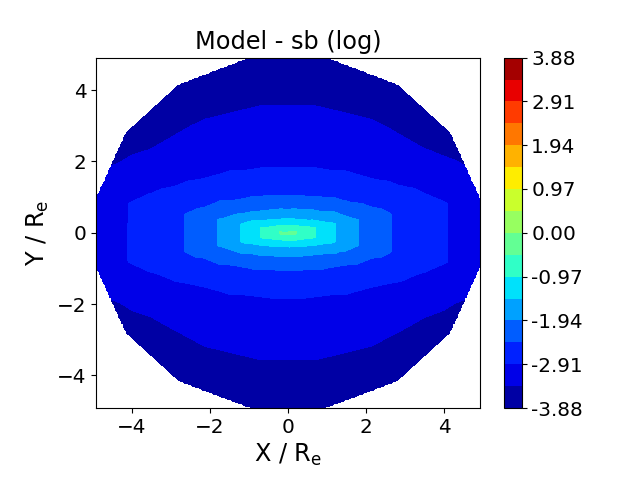}\\
		\includegraphics[width=45mm]{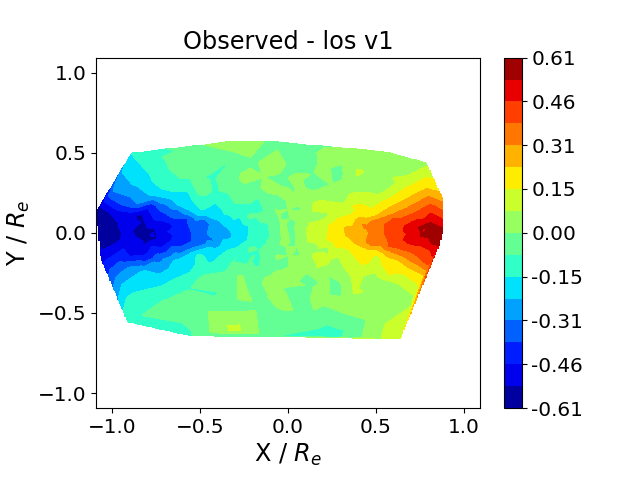}  & \includegraphics[width=45mm]{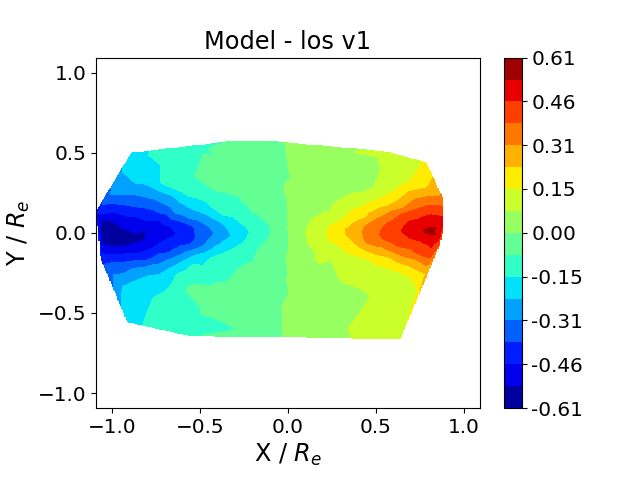}  & \includegraphics[width=45mm]{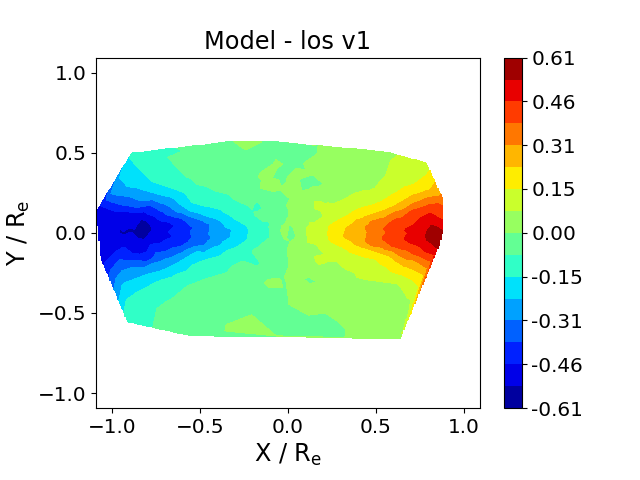}\\
		\includegraphics[width=45mm]{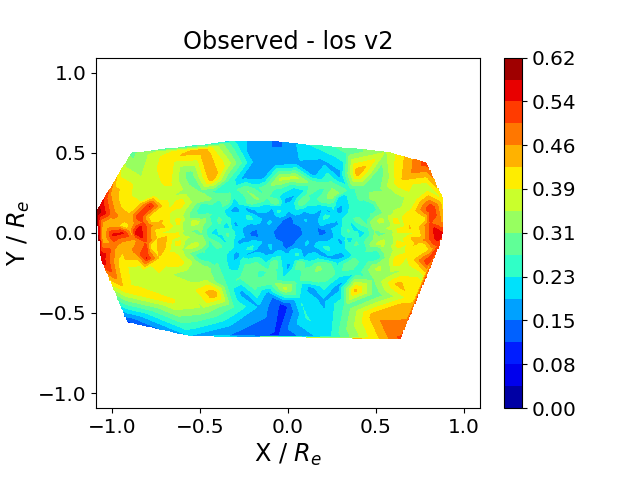}  & \includegraphics[width=45mm]{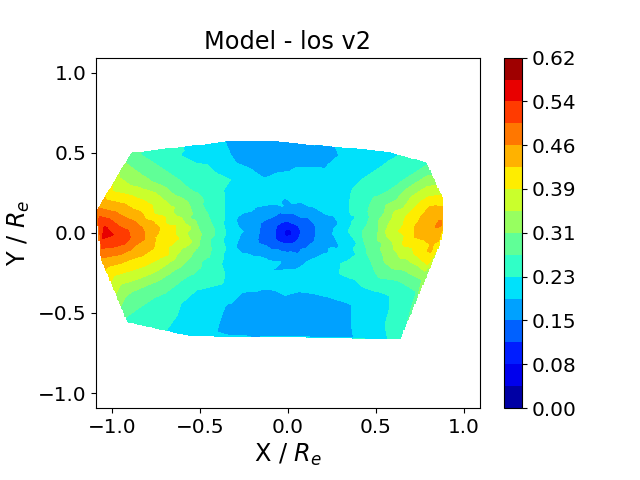}  & \includegraphics[width=45mm]{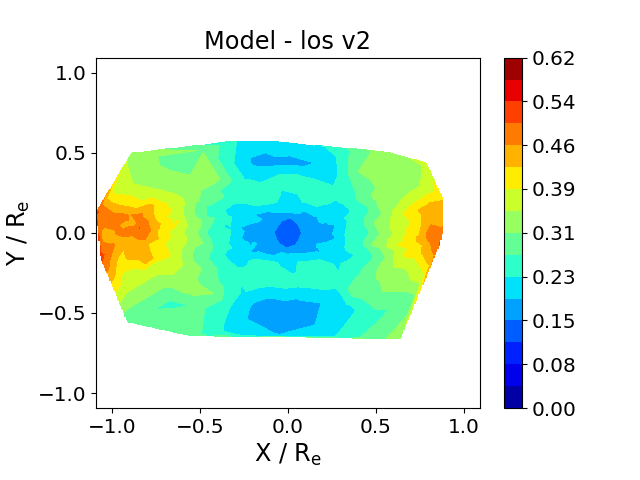}\\
	\end{tabular}
	
	\medskip
Reproduction of observables by M2M and SCHW using NGC 4452 with 3I initial conditions (Appendix \ref{app:obsrepro}). The observables are surface brightness in the top row followed by the first and second velocity moments.  Units are as in Section \ref{sec:galdata}. The first column displays the observed data used in modeling, with the second and third columns showing reproduction of the observed data by the modeling methods.  In row order, the mean observable $\chi^2$ values are ($0.76$, $0.42$, $0.69$) for M2M and ($0.30$, $0.13$, $0.25$) for SCHW (see Table \ref{tab:AtlasMSchi2}).
\end{figure}

This appendix contains a visual example in Figure \ref{fig:obsrepro4452} of the reproduction of observables by M2M and SCHW using NGC 4452 with 3I initial conditions (see Section \ref{sec:M2MSCHW3I}).

\end{document}